# Photoelectron angular distributions for photoionization of argon by two-color fields

**Soumia Chqondi[1, 2,*], Souhaila Chaddou[2] and Abdelkader Makhoute[2]**

[1]*Laboratory ISTM (Innovation in Sciences, Technologies, and Modeling) Physics Department,*

*Faculty of Science. Chouaïb Doukkali University El Jadida, Morocco*
*(Email: chqondi.s@ucd.ac.ma)*

[2]*Physics of Radiation and Laser-Matter Interactions, Faculty of Sciences,*
*University of MoulayIsmail, Meknès, Morocco*

## Abstract

We perform numerical simulations for photoionization of the argon atom when exposed to a combined field of an infrared laser (abbreviated IR) and its 13th harmonic (H13) by solving the Time-Dependent Schrödinger Equation (TDSE)], noting that the associated energy spectrum displays a harmonic peak dressed by the IR and other peaks called "side-bands" SB±n corresponds to the absorption (+n) and the emission (-n) of the n-infrared photons above the ionization threshold. In two-color H13+IR photoionization of argon, the analysis of the angular distribution of the ejected electron in the continuum, and our calculation approach is based on the method of projection of the electronic wave function on the states of the continuum, the latter was found in using the Runge-Kutta 4 method and the Schutting method. Our simulation results demonstrate that either for the harmonic peak or the sideband, the shape of the angular distribution exhibits a good agreement with the interpretation based on the generalized Fano rule for the absorption and emission of photons.

**\*Corresponding author**

chqondi.s@ucd.ac.ma

**Faculty of Sciences University Chouaib Doukkali, B. P. 20, 2300, El Jadida, Morocco.**

Adresse: Route Ben Maachou.

Tél. : 05.23.34.23.25 / 05.23.34.30.03. Fax: 05.23.34.21.87.

# Introduction

The combination of two pulses instead of just one, in very different wavelength regimes, has demonstrated that it is an ideal tool to explore different aspects during the interaction of an atom with an intense field composed of two fields.

Several theoretical and experimental studies [1, 2,3] have focused on the mechanisms involved in the process of ionization by absorption of two photons. The theoretical study of these processes is complex. Indeed, the theoretical developments used until then, such as the theory of perturbations, become obsolete. Another theoretical approach has thus been worked out to obtain the atomic answer, it is based on the resolution of the time-dependent Schrödinger equation (TDSE), a fundamental equation of quantum physics. For multi-electron atoms, such as Ne and Ar, due to their electronic structure, a direct numerical solution of Schrodinger's equation is extremely difficult, and only approximations have been used so far. In many cases, solving ESDT is done in a simple quasi-static approach where the atomic system is treated by a single-active-electron (SAE) model [4, 5], ignoring the coupling of this active electron with other electrons and the influence of external field on them, i.e. only the most weakly bound electron (a valence electron) interacts significantly with the laser field, its interaction with the electronic nucleus is defined by a model potential well adapted to the atomic system.

In this paper, we perform the IR laser-assisted EUV photoionization simulations of argon by numerically solving the TDSE, describing the evolution over time of the wave function of the electrons interacting with an intense laser field, in the framework of the single-active-electron approximation (SAE). We focused on describing the process of non-resonant two-photon ionization of argon, where we used an IR laser delivered by the Ti. We have chosen the 13th harmonic (noted H13), in the EUV range, to directly ionize the argon atom from a state whose outermost "active" electron is in the ground state of the 3p orbital, so it has p symmetry which can initially be in the pσ (m = 0) and pπ (m = 1) states. Due to the axial symmetry of the problem, the ionization of the symmetry states σ (m = 0) and π (m = 1) can be considered independently.

The associated energy spectrum then shows harmonic peaks dressed by IR and other peaks called "side-bands" SB±n (Fig. 1), which are the signature of H13+IR two-photon transitions. The allowed multiphoton transitions are obtained by applying the angular momentum selection rules where the orbital angular quantum number l must change by one unit $\Delta \ell = \pm 1$. In addition to the selection rules, Fano proposed a propensity rule [6], which states that among the two possible transitions, the one which increases the angular momentum of the electrons, $\ell_{max} \rightarrow \ell+1$, is favored over the one which decreases the angular momentum, $\ell_{max} \rightarrow \ell -1$, due to the increase in potential centrifugal with the angular momentum in the case of absorption, and the reverse in the case of emission. This rule is then generalized to laser-assisted photoionization [7], consisting of the absorption of a harmonic photon (in the extreme ultraviolet), associated with the exchange of two IR photons by absorption or stimulated emission.

Our simulation results showing the angular distribution of photoelectrons as a function of the emission angle must respect the selection rule. The possible transitions during the absorption of a high-frequency harmonic photon from the ground state of argon 3p ($\ell = 3$, m=0, |m| = 1) will be p→s ($\ell$ =0) or p→d ($\ell$ =2) with the dominance of d-wave according to Fano's rule stated above. In the continuum the ionized electron can absorb or emit IR photons above-threshold ionization, the possible transitions are then d→p ($\ell$ =1) or d→f ($\ell$ =3) with the dominance of the f-wave for absorption and p-wave for emission according to the generalized Fano's rule.

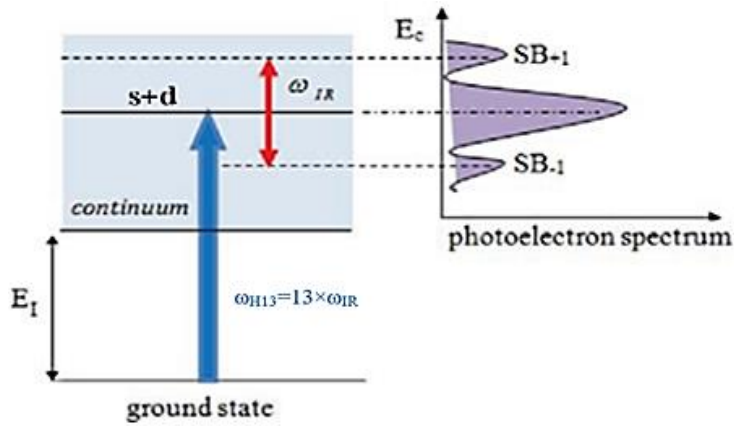

FIG. 1: *An illustration of the photoelectron spectrum obtained during two-color photoionization. The atom that has been ionized by a harmonic photon (blue flash) absorbs or emits IR photons above-threshold ionization (red flash).*

**Theoretical and Numerical Methods**

**Laser Field Characteristics :**

The interaction field we used results from the combination of an IR laser field delivered by a Ti: Saphir with a frequency $\omega_{IR}$ = 0.057 a.u. ≡ 1.55 eV and one of its frequency harmonics $\omega_{Hq}$ = q × $\omega$IR with q = 2a + 1 (a being an integer). The value of q must be large enough for the atom to be ionized by a single photon (q=13 in our case). Both fields have the same linear polarization and the same total duration. The representation of the resulting electric field can be in the following form

with
$$\begin{cases} \vec{E}_T(\vec{r},t) = \vec{E}_{Hq}(\vec{r},t) + \vec{E}_{IR}(\vec{r},t) \\ \vec{E}_{Hq}(\vec{r},t) = E_{0Hq}f(t,\omega_{IR})\sin(\omega_{Hq}t)\vec{e}_z \\ \vec{E}_{IR}(\vec{r},t) = E_{0IR}f(t,\omega_{IR})\sin(\omega_{IR}t)\vec{e}_z \end{cases} \quad (1)$$

For a sinusoidal envelope, the function $f(t, \omega_{IR})$ takes the following form :

$$f(t, \omega_{IR}) = \sin^2\left(\frac{\pi t}{t_{max}}\right) \quad (2)$$

The total pulse duration Tmax is defined from the period of the infrared laser $T_{IR}$ by

$$T_{max} = n_{c.o} T_{IR} = n_{c.o} \frac{2\pi}{\omega_{IR}}, \quad (3)$$

where $n_{c.o}$ is the number of optical cycles of the laser field.

**Time-Dependent Schrödinger Equation**

The time-dependent Schrödinger equation describing the laser-hydrogen interaction is written in the dipole approximation and a.u. in the following form

$$i\frac{\partial}{\partial t}\psi(\hat{r}, t) = H(\hat{r}, t)\psi(\hat{r}, t), \quad (4)$$

where $H$ is the total Hamiltonian of the system, which is decomposed as follows

$$H = H_0 + H_{int}(t), \quad (5)$$

$H_{int}$ is the interaction Hamiltonian representing the interaction between the electron and the electric fields given by $\vec{E}(t).\vec{r}$ in the gauge of length.

$H_0$ is the Hamiltonian of the atom

$$H_0 = -\frac{1}{2}\nabla^2 + V(r) \quad (6)$$

The model potential well suited to argon in the single active electron approximation (SAE) takes the following Klapisch form [8]

$$V(r) = -\frac{1+(Z-1)e^{-a_1 r}+a_2 r e^{-a_3 r}}{r} \quad (7)$$

This type of potential satisfies the boundary conditions

$$\begin{cases} \lim_{r \to 0} V(r) = -\frac{Z}{r} \\ \lim_{r \to \infty} V(r) = -\frac{1}{r} \end{cases} \quad (8)$$

where Z represents the nuclear charge. The $\{a_i\}$ parameters are adjusted to have energy values close to the experimental values. The eigenstates ψn and the corresponding eigenvalues En are numerically evaluated by the diagonalization in the spherical coordinates of the Hamiltonian H0 carried out by a program developed by our team and which is based on the QL method [9, 10].

The numerical values obtained by the numerical resolution of the Schrödinger equation were compared with those of reference at the NIST (National Institute of Standards and Technology) [11].

To have numerically processed our laser-argon atom interaction process, we integrate the resulting ESDT (4) in time using the Peaceman-Rachford algorithm [12] also called the alternating directions method (ADI, Alternate Direction Implicit). This stable method exploits the fact that, in practice, the tridiagonal matrices of H0 and Hint can easily be reversed.

## **Angular distribution**

After having presented the energy spectrum of the ejected electron, it would be interesting to have the angular distribution of this electron concerning the direction of polarization, which will inform us about the mechanisms involved. The calculation of the distribution angular has the interest of testing the different analytical or numerical approaches, which require the atomic answer to intense laser fields. Different methods exist to calculate the angular distributions of photoelectrons produced during two-color photoionization, such as the one based on the calculation of the flux of photoelectrons in a given energy window [13]. The chosen method is very simple and is based on the projection of the final wave function $\psi(\hat{r}, t_{max})$ obtained by solving the TDSE at the end of the interaction, on the discretized continuum eigenstates $\psi_{ck}(\hat{r})$ associated with the energy $E_k = \frac{k^2}{2}$ such that the differential ejection cross-section in the solid angle dΩ is

$$\left(\frac{d\sigma}{d\Omega}\right)_{E_k} \propto |\langle \psi_{ck}(\hat{r}) | \psi(\hat{r}, t_{max}) \rangle|^2 \qquad (9)$$

Taking into account the cylindrical symmetry of the problem, the solid angle reduces to the angle theta between the wave vector of the photoelectron $\tilde{k}$ and the polarization direction of the fields that we have chosen parallel to Oz.

The wave function ψ(r̂,t) can be written as a sum over the radial functions $u_\ell(r, t_{max})$ times the spherical harmonics Yl,m

$$\psi(r, \theta, \varphi, t_{max}) = \sum_{\ell=0}^{\ell_{max}} \sum_{m=-\ell}^{\ell} \frac{u_\ell(r, t_{max})}{r} Y_{\ell,m}(\theta, \varphi) \qquad (10)$$

The expression of the continuum wave function $\psi_{ck}(\hat{r})$ is given by [14]

$$\psi_{ck}(\hat{r}) = \sum_{\ell=0}^{\ell_{max}} \sum_{m=-\ell}^{\ell} Y_{\ell,m}^*(\widehat{K}) Y_{\ell,m}(\theta, \varphi) i^\ell e^{-i(\sigma_\ell + \delta_\ell)} R_{k,\ell}(r) \qquad (11)$$

with $\sigma_\ell = \text{Arg}(\Gamma(\ell + 1 - \frac{i}{k}))$ is the Coulomb phase, $\delta_\ell$ is the non-Coulomb phase, and $\mathbf{R_{k,\ell}(r)}$ is the set of eigenstates of the continuum.

The differential ejection cross-section is

$$\left(\frac{d\sigma}{d\theta}\right)_{E_k} \propto \left| \sum_{\ell,\ell'} \sum_{m,m'} (-i)^\ell e^{(\sigma_\ell + \delta_\ell)} \langle Y_{\ell,m} | Y_{\ell',m'} \rangle \langle R_{k,\ell} | u_{\ell'} \rangle Y_{\ell,m}(\widehat{K}) \right|^2 \qquad (12)$$

Knowing that $\langle \mathbf{Y_{\ell,m}} | \mathbf{Y_{\ell',m'}} \rangle = \boldsymbol{\delta_{\ell,\ell'} \delta_{m,m'}}$ and setting $S_\ell = \langle \mathbf{R_{k,\ell}} | \mathbf{u_\ell} \rangle$, the above quantity can be expressed more simply in the form

$$\left(\frac{d\sigma}{d\theta}\right)_{E_k} \propto \left| \sum_\ell \sum_m (-i)^\ell e^{(\sigma_\ell + \delta_\ell)} S_\ell Y_{\ell,m}(\widehat{K}) \right|^2 \qquad (13)$$

The angular distribution of the peak, whether harmonic or sideband, is only the sum of the angular distributions over the energies corresponding to the points of this peak.

$$\left(\frac{d\sigma}{d\theta}\right)_{peak} = \Sigma_{E_k} \left(\frac{d\sigma}{d\theta}\right)_{E_k} \quad (14)$$

To approach the problem numerically, it is necessary to first find the functions of the continuum Rk. To do this, we can numerically solve the stationary Schrödinger equation (ESS) H0Rk, (r) = EkRk, (r).

The process employed to extract the electronic eigen-energies from the continuum is called "shooting". The basic concept is to pull a point on the photoelectron spectrum that is connected to very precise energy, such as $E_k = \frac{k^2}{2}$, and then search from that point for the corresponding discrete state. For the second step, we decided to use a numerical program for the integration of differential equations, it is fourth-order Runge–Kutta algorithm.

## **RESULTS AND DISCUSSION**

As mentioned above, the system chosen for the numerical simulation of the two-color photoionization process is that of an argon atom subjected to pulses combining IR photons from the Ti: sapphire laser of frequency ωIR = 0.057 a.u. = 1.55eV and its 13th harmonic H13 of frequency ωH13 = 13 × ωIR = 0.741 a.u≡20.155 eV, is well within the UV wavelength range (à verifier).

If an argon atom is subjected only to the harmonic field H13 for a weak intensity than that of the generator laser, it can ionize and release an electron whose kinetic energy, determined by the energy conservation law, is the difference between the energy of the absorbed photons and the ionization potential of argon

$$E_c(H13) = \omega_{H13} - I_P \quad (15)$$

where the ionization potential of the argon atom $I_P$ is 0.579 a.u.

When the 13th harmonic pulse of the IR laser is superimposed on the pulse of the fundamental ("dressing") laser, new peaks in the photoionization spectrum in the form of sideband peaks designated SB±n arise around the harmonic peak H13. These sideband peaks correspond to the exchange of n IR photons (with n >=1) via absorption or stimulated emission. The ejected electron acquires in the continuum kinetic energy, which will move to the left by a quantity Up, the energy conservation law is rewritten as follows :

$$E_c(H13) = \omega_{H13} - (I_P + U_P), \quad (Harmonic\ peak) \quad (16)$$

or $\quad E_c(SB_{\pm n}) = \omega_{H13} \pm n\omega_{IR} - (I_P + U_P), \quad (Sidebands) \quad (17)$

where $U_P(\text{a.u.}) = \frac{E_0^2}{4\omega_{IR}^2}$ is the ponderomotive potential corresponding to the average energy of vibration of the electron in the laser field, $U_P$ then depend on the intensity and the frequency of the dressing field.

To perform the numerical simulation of the two-color photoionization process of our system, we must first determine the harmonic intensity and the intensity of the fundamental laser radiation.

In the absence of an IR laser field, the argon atom is only exposed to a harmonic field H13 of frequency $\omega_{H13}$ selected high enough to ionize the atom by a single photon ; the photoelectron spectrum shows a single peak attributed to the absorption of a harmonic photon (see Fig. 2). According to this figure, we note the appearance of a harmonic peak linked to the direct ionization by a single harmonic photon (EUV photon) located at an energy $E_c(H13) = 0.162\ a.u.$, which is in good agreement with the results calculated by the energy conservation law defined by the equation (15). We also notice e that the probability of ionization depends on the magnetic quantum number, it is more pronounced for m = 0 than for m = 1. These results demonstrate that only the m=0 component of the initial atomic state can be chosen.

In the continuum, the photoelectron wave function resulting from the 13th harmonic ionization from the 3p ($\ell = 1$) ground state must have either s-wave ($\ell = 0$) or d-wave ($\ell = 2$) character due to the selection rule. The shape of the angular distribution of the photoelectrons, as shown in Fig. 3, is extremely similar to a d0-wave with m = 0 or a d1-wave with m = 1. This demonstrates that in the case of absorption, the angular momentum of the electrons $\ell = 2$ is favored over $\ell = 0$. Our simulation results are consistent with the propensity rule presented in the introduction.

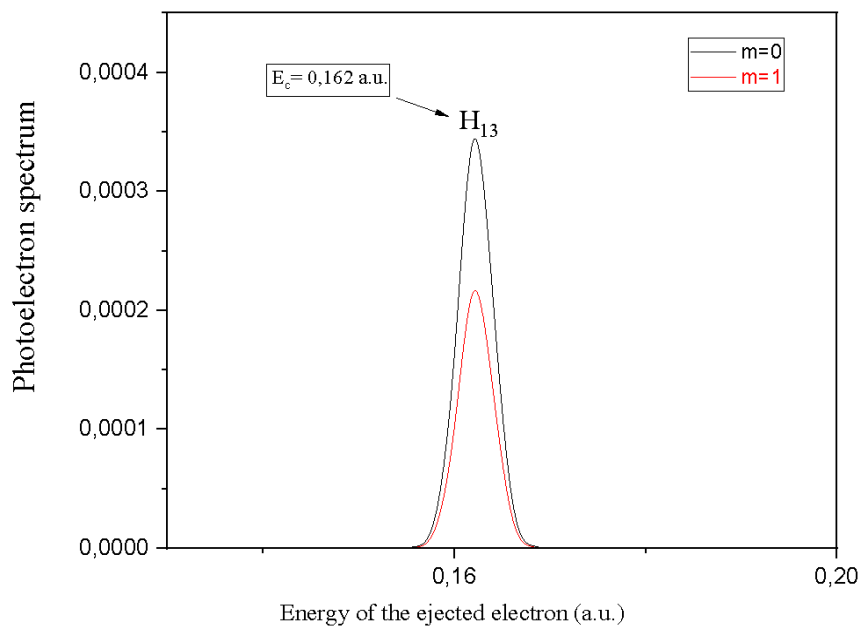

FIG. 2 : *Photoelectron energy spectrum of the direct photoionization of argon (3p) by absorption of the photon H13 at the quantum numbers m=0 (black curve) and dm=1 (red curve).*

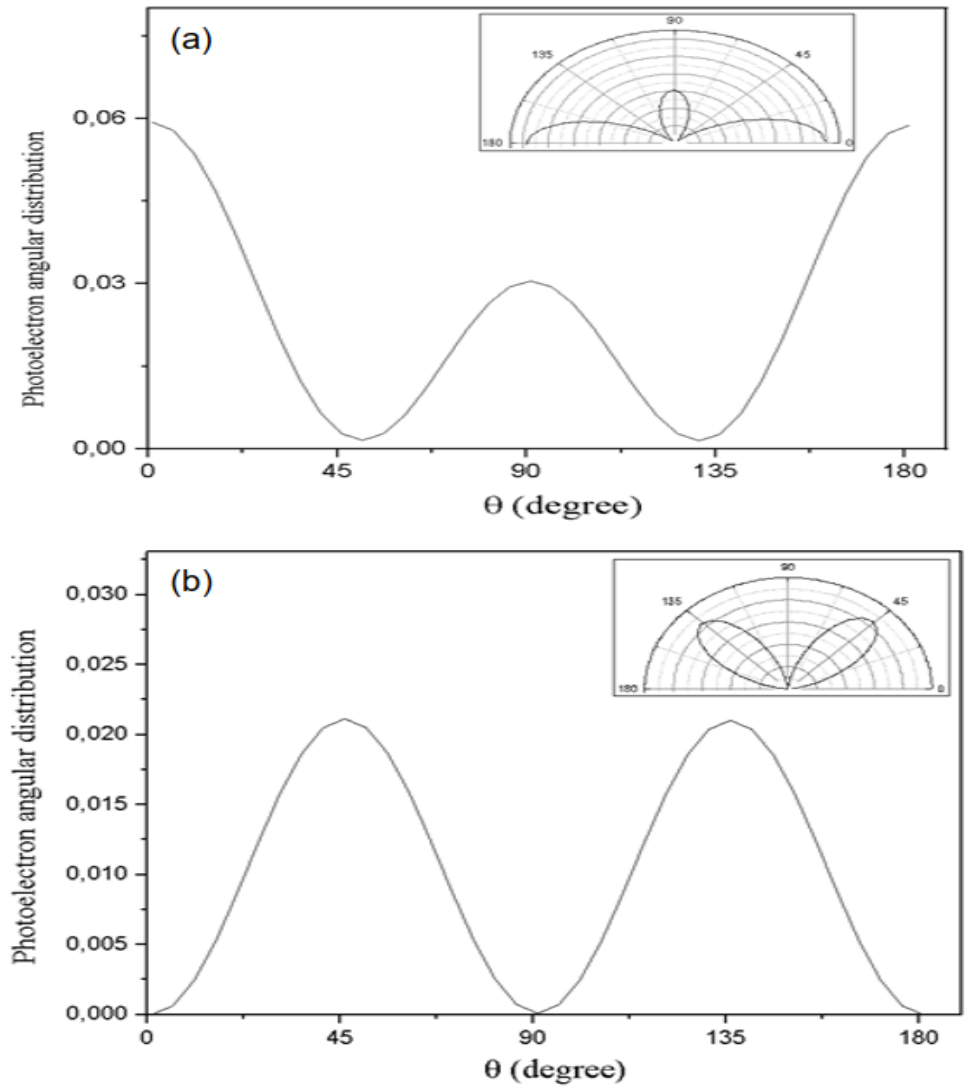

FIG. 3: *Photoelectron angular distribution of H13 as a function of the angle θ for the quantum numbers m=0 (a) and m=1 (b).*

Maintaining the same harmonic intensity and superimposing it on an infrared field, As indicated in the introduction, new sideband peaks $SB_{+1}$ and $SB_{-1}$ then appear around the harmonic peaks which correspond respectively to the absorption or the emission of n infrared photons (see Fig. 4).

Figure 4 shows that with even greater infrared illumination, it will be possible to see higher-order sidebands ($SB_{+2}$, $SB_{-2}$, etc.), such that their amplitude increases while that of the harmonic peak decreases, everything happens as if the harmonic peak were partly transformed into satellite peaks.

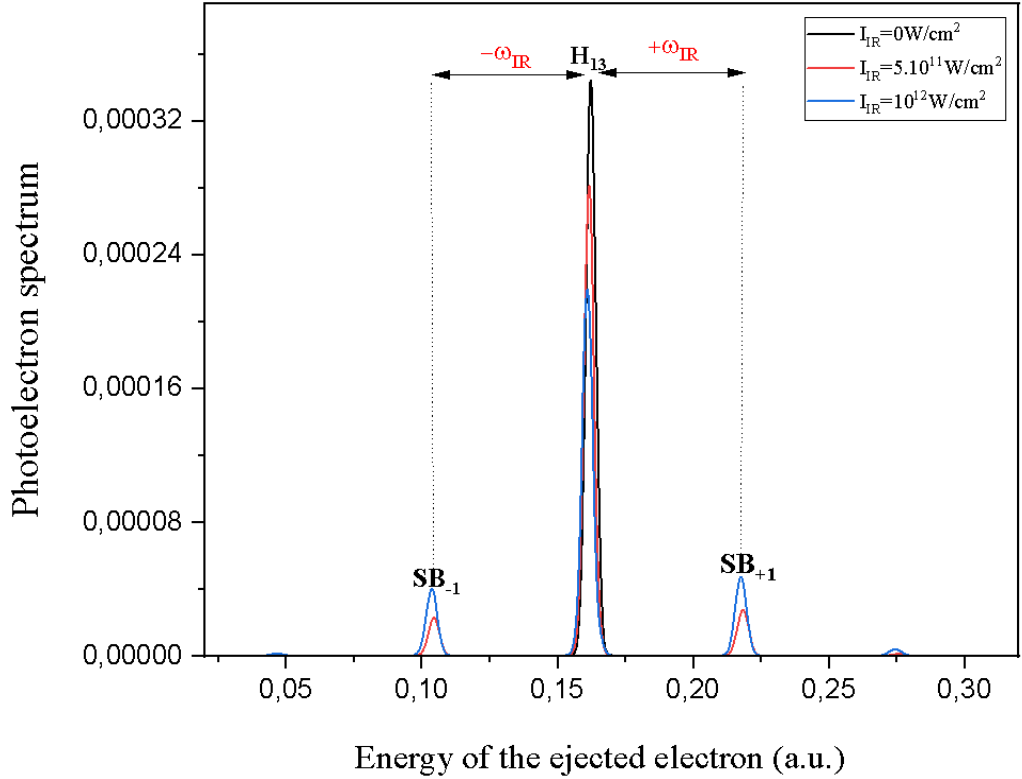

FIG. 4: *Photoelectron energy spectrum of argon for a harmonic intensity IH13 =$10^{10}$W/cm² and different infrared intensities*

Taking the case where the infrared intensity is *$5.10^{11}$W/cm²*, the ponderomotive energy Up, which causes the displacement of the ionization threshold, for this IR intensity is $U_P = 1.1 \times a.u.$. According to Fig. 4, we note the appearance of a harmonic peak linked to the absorption of the harmonic photon *H13* located at an energy $E_c(H13) = 0.161\, a.u.$, surrounded by two sideband peaks $SB_{+1}$ and $SB_{-1}$ placed symmetrically from each side and separated by the same amount of energy ωIR, they are associated respectively with the absorption and the stimulated emission of IR photon in the continuum. These sideband peaks are positioned at energies $E_c(SB_{+1}) = 0.218\, a.u.$ and $E_c(SB_{-1}) = 0.104\, a.u.$. These results are in good agreement with those calculated by the energy conservation law established by equations (16) and (17).

As we mentioned before, the possible transitions for the absorption ($SB_{+1}$) or the emission ($SB_{-1}$) of the infrared photon in the continuum are $d \to P(\ell = 1)$ or $d \to f(\ell = 3)$.

We can see from Fig. 5 that the angular distributions of the $SB_{+1}$ and $SB_{-1}$ peaks exhibit different behaviors, although they both exhibit a minimum for m=0 and a maximum for m=1 in $\theta = 90°$. Figs. 5 (a) and (b) show the shapes of the angular distributions of the $SB_{+1}$ peak, which are similar to that of the wave $f_\sigma$ (m=0) and $f_\pi$ (m=1), respectively. Indeed, its evolution as a function of theta shows three minima and four maximum for m=0, whereas the number of minima and maxima is reversed for m=1. While the angular distributions of the SB$_{-1}$ peak in Figs. 5(c) and (d) have shapes very comparable to those of the p$_\sigma$ (m=0) and p$_\pi$ (m=1) waves, respectively, where we find a minimum and two maxima for m=0, and the reverse for m=1, that is to say, two minima and a maximum.

Our methods for calculating the angular distributions of photoelectrons in the case of the two-color H13+IR photoionization of the argon atom have been validated thanks to our results of simulations found, which allow us to verify the generalized Fano rule.

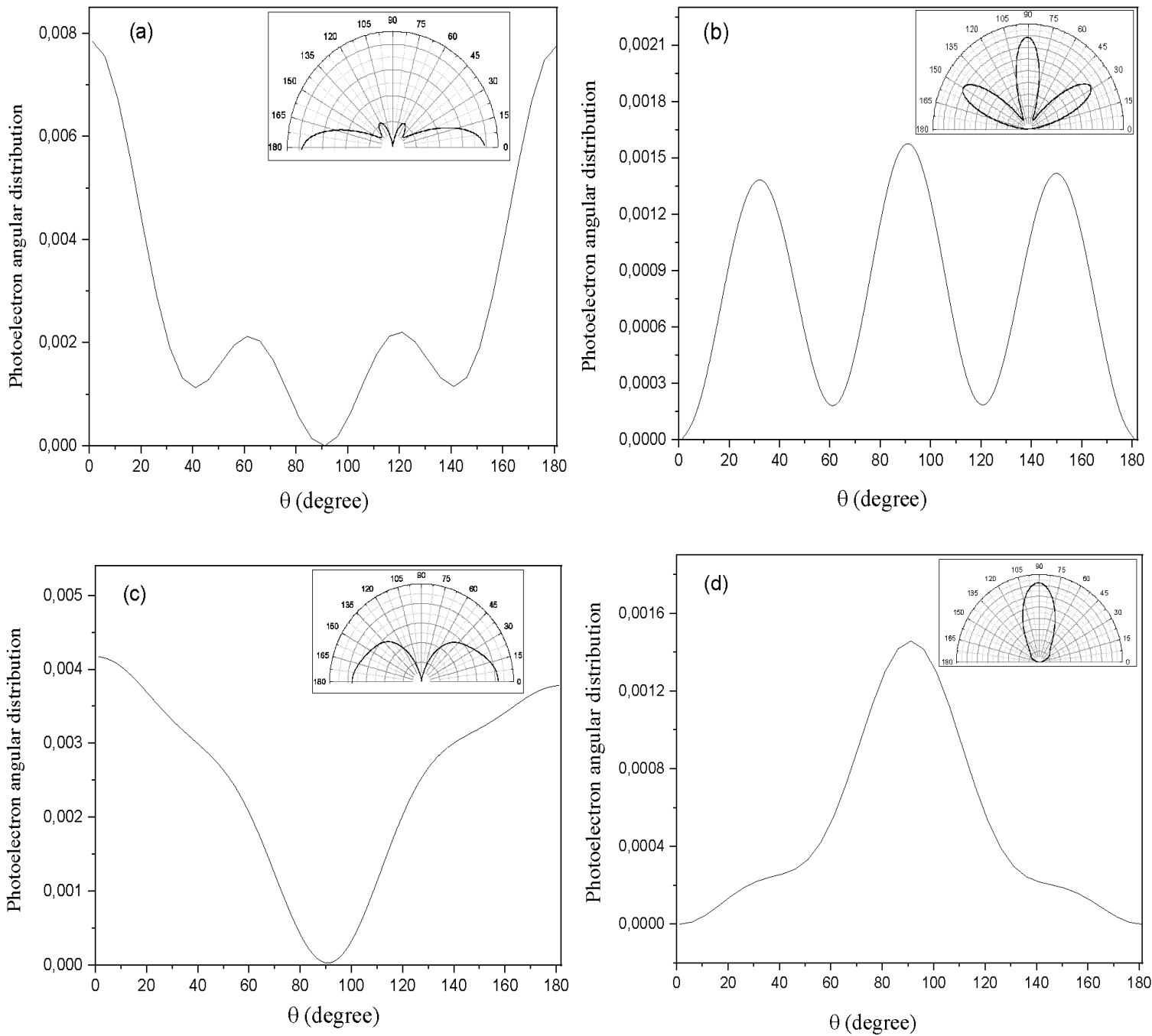

FIG. 5: *Angular distribution of photoelectrons as a function of the angle θ of the sideband SB$_{+1}$ for m=0 (a) and m=1 (b) and of the sideband SB$_{-1}$ for m=0 (c) and m= 1 (d).*

# Conclusion

In this paper, we studied the angular distribution of the electron ejected by the two-color H13+IR photoionization of the argon atom and chose to use the thirteenth harmonic of the infrared laser to move away from the Cooper minimum in argon. The allowed electronic transitions are obtained by applying the selection rule for the orbital quantum number $\Delta \ell = \pm 1$ mentioned above, which then gives two possibilities of transitions, the one that must be brought into play is given by the rule of Fano.

We have presented the energy spectrum of the photoelectrons, noting an excellent agreement with the law of conservation of energy in terms of localization of kinetic energy. The shape of the angular distribution of the peaks of the spectrum represents an alternation of the maximum and the minimum, which fits the shape of the spherical harmonics $Y_{\ell+1}$, m for the absorption and $Y_{\ell-1}$, m for emission, these results are those predicted by applying Fano's rule, thus confirming the quality of the methods used in our calculations of the angular distribution of the argon atom.